\documentclass[12pt]{spieman}  
\usepackage{amsmath,amsfonts,amssymb}
\usepackage{graphicx}
\usepackage{setspace}
\usepackage{tocloft}
\usepackage{soul}
\usepackage[left]{lineno}
\usepackage{comment}

\title{A bestiary of low-level electrical artifacts in SPHEREx flight data}

\author[a,*]{Candice M. Fazar}
\author[a]{Michael Zemcov}
\author[b]{C. Darren Dowell}
\author[b]{Brendan P. Crill}
\author[c]{Chi Nguyen}
\author[c,b]{Howard Hui}
\affil[a]{Rochester Institute of Technology, Center for Detectors, College of Science, 1 Lomb Memorial Drive, Rochester, NY, 14623}
\affil[b]{Jet Propulsion Laboratory, California Institute of Technology, 4800 Oak Grove Drive, Pasadena, CA 91011}
\affil[c]{California Institute of Technology, 1200 E California Blvd,  Pasadena, CA 91125}

\cftpagenumbersoff{figure}
\cftpagenumbersoff{table} 
\begin{document} 
\maketitle

\begin{abstract}
SPHEREx, NASA’s all-sky spectrophotometric survey mission, maps the sky in 102 spectral channels from 0.75~$\mu$m to 5.0~$\mu$m at 6.15$^{\prime \prime}$ resolution from low-Earth orbit using a custom readout of six HAWAII-2RG detectors. The SPHEREx instrument team and Science Data Center have built a robust science data pipeline that includes mitigating image-space artifacts arising from low-level optical and electronic effects. Analysis of first-survey flight data has confirmed predictions for some of these effects, determined differences between the laboratory and flight predictions in others, and even found new artifacts that were unanticipated. This paper will summarize our current understanding of artifacts in SPHEREx images, including image-space artifacts resulting from cosmic rays, crosstalk and unresponsive pixels. We also present mitigation and/or masking strategies implemented to mitigate such effects, and an assessment of the possible implications for other space missions using HAWAII-2RG detectors.
\end{abstract}


{\noindent \footnotesize\textbf{*}Candice M. Fazar,  \linkable{cmfsps@rit.edu} }

\section{Introduction}
\label{s:intro}  

HgCdTe HAWAII-2RG hybridized detectors (H2RGs) are the detector arrays of choice for low background infrared astronomical observations.  Having been developed for JWST~\cite{Garnett2004}, Teledyne 2.5~$\mu$m and 5.0~$\mu$m cutoff HgCdTe detectors hybridized to the HAWAII-2RG multiplexer exhibiting low noise and low dark current are commercially available and have been used in space-based astronomy missions such as CIBER, CIBER-2, JWST, EUCLID, and SPHEREx, amongst others~\cite{Garnett2004, Zemcov2010, Zemcov_2025, Crouzet2012, Bock_2026}.  Each mission utilizes read-out strategies optimized for its science goals, leading to unique systematic anomalies that must be mitigated.  Those that manifest as 
image-space artifacts can be particularly challenging for science requiring high fidelity measurements of diffuse emission, such as the EBL science goal of the SPHEREx mission~\cite{Cheng_2022, cukierman2026spectralmapmakingspherex}.

SPHEREx is an all-sky spectral survey mission, obtaining spectral maps in 102 bands from 0.75 to 5.0~$\mu$m.  It launched on 11 March 2025 and as of the writing of this paper has completed one full year of its survey plan, including two all-sky spectral maps.  The telescope focal plane is populated with six H2RGs: three 2.65~$\mu$m cutoff H2RG detectors forming the short-wave infrared (SWIR) focal plane assembly (FPA) and three 5.4~$\mu$m cutoff H2RG detectors forming the mid-wave infrared (MWIR) FPA, each overlaid with a linear variable filter (LVF) defining its wavelength bands.  While the LVF captures spectral features of astrophysical targets, it also records foregrounds, such as airglow~\cite{hui2026observationsatmosphericheliumoxygen} in wide spectral bands.  Similarly, the detectors themselves not only record signal from distant objects, it also records the signal generated when cosmic rays and other particles encounter the H2RG detectors as it proceeds in its orbit.\cite{Nguyen2026} These can manifest as image-space artifacts, particularly if the charge deposited is insufficient to trigger transient detection. The data processing pipeline, which processes downlinked data into calibrated images for science analysis, includes modules to flag such features.  

Many image-space artifacts both electronic (i.e.~image persistence, crosstalk and $1/f$ noise) and optical (i.e.~dichroic beam splitter reflections) have been characterized in the laboratory~\cite{korngut2026spherexinstrumentcalibrationtesting, Fazar2025} and have undergone or are now undergoing an updated characterization campaign in flight.  Other effects were not well characterized in the lab due to the difference between laboratory and flight conditions.  In particular, the impacts of energetic cosmic rays and high flux sources coupled with the read-out scheme were not tested in the lab, leading to additional forms of crosstalk and an extended measured PSF for very bright sources observed in flight.  With the wealth of flight data and repetitiveness of the read-out strategy, a variety of expected and new image-space artifacts are being cataloged and flagged within the L2 pipeline.  Here we present our current understanding, based upon one year of L1 and L2 data, of the unanticipated electronic artifacts in SPHEREx images, including how they manifest, the steps we have taken to characterize them, and mitigation strategies, such as masking.

\section{Data}
\label{s:data}

SPHEREx detectors operate in sample-up-the-ramp (SUR) mode, meaning that their signal is sampled non-destructively as the charge accumulates.  Each sample of the full detector array corresponds to a single frame, of which there are currently 77 in the SUR integration.  The unique read-out schemes for noise suppression~\cite{Nguyen2025, Heaton_2023} utilized by SPHEREx to accomplish its science goals result in non-sequential row reads and additional samples of reference (non-photosensitive border) pixels and Video 8 (detector readout amplifier) voltages (``phantom" pixels) with each frame.  Due to data volume limitations, per-pixel slopes and offsets are calculated on board the spacecraft,\cite{Zemcov_2016} and slopes and associated SUR bit flags are downlinked in Level-0 (L0) data streams.  Raw images and existing flag bits are then assembled in detector order on the ground, corrected for electronic drift using reference and ``phantom" pixels and converted from Analog to Digital Units (ADU) to $e^-/$s, amongst other processing and packaging steps, resulting in L1 data.~\cite{explanatory_supplement}  
Finally, these L1 images are processed into L2 images on the ground, a process which includes astrometric and photometric calibration;  
see the SPHEREx Explanatory Supplement~\cite{explanatory_supplement} for details.

The bit mask flag layer in L1 data reports any anomalies detected in the determination of the slope~\cite{Zemcov_2016} with three basic SUR flags generated onboard the spacecraft, two of which are TRANSIENT and OVERFLOW, identifying transient detection and the onset of $\sim5\%$ non-linearity respectively.  
Both of these conditions terminate the SUR slope-calculation process, which starts with the third sample after detector reset.  In both cases, the slope identified prior to this triggering event is recorded if a sufficient number of samples 
are present to estimate a slope.  Otherwise, the slope is assigned NaN in the flux image.
The third basic SUR flag is SUR\_ERROR, which, when paired with TRANSIENT or OVERFLOW, is designed to signify statistically different noise properties or errors in the SUR measurement.  This flag is triggered with TRANSIENT if the reported slope is determined using fewer than TCUT frames, currently set at 38, which corresponds to approximately half of the full SUR data. Consequently, the SUR\_ERROR flag divides the TRANSIENT group into early and late transients occurring in the first and last half of the exposure respectively.  
This same flag is triggered with OVERFLOW when there are insufficient data points to obtain a slope fit, identifying early overflow pixels typically subjected to heavy flux.  The combination of all three basic SUR flags with a measured flux indicates that the transient was of sufficient energy to trigger the overflow flag and that the event occurred in the first half of the exposure.  Likewise, TRANSIENT with OVERFLOW alone indicates a high energy transient in the latter half of the exposure.  Representative of various circumstances, and tabulated as combinations by the integer sum of their SUR flag bits for quick reference in Table~\ref{tab:flags}, 
these flags are used in conjunction with the slope-fit image to identify pixels that should be discarded from science analyses based on features identifiable in the data time series. 

\begin{table}[h]
    \centering
    \begin{tabular}{l|ccc|c}
         & TRANSIENT & OVERFLOW & SUR\_ERROR &  \\
        \hline \hline
        CATEGORY & BIT 0 & BIT 1 & BIT 2 & INTEGER SUM \\
        \hline \hline
        Late Transient  & 1 & 0 & 0 & 1 \\
        Overflow & 0 & 1 & 0 & 2 \\
        Late Overflow-Transient & 1 & 1 & 0 & 3 \\
        Early Transient & 1 & 0 & 1 & 5 \\
        Early Overflow & 0 & 1 & 1 & 6 \\
        Early Overflow-Transient & 1 & 1 & 1 & 7 \\
    \end{tabular}
    \vspace{0.1cm}
    \caption{Combinations of basic SUR flags relevant to Sections \ref{s:blooming} and~\ref{s:snowballs} are connected here to their integer sum, which is used to illustrate representative groupings of pixels in Figs \ref{fig:bloomexamples} and~\ref{fig:snowexamples}.}
    \label{tab:flags}
\end{table}

In addition to these basic SUR flags, the L1 flag layer includes bits representing known features of the imaging system, such as bad pixels and regions of low optical efficiency.  During L2 processing, flag bits representing observation-dependent image-space features are added, including 
flags for known sources, image persistence from a prior observation, and ghost images due to stray light, amongst others.  In the following sections, we discuss three categories of unanticipated electronic artifacts now flagged in the L2 processing pipeline, including charge blooming, high-energy cosmic ray and snowball impacts, and three types of crosstalk.

\section{Charge Blooming}
\label{s:blooming}

In analyzing L1 data, it was discovered that nearly every source bright enough to trigger the OVERFLOW flag also triggered TRANSIENT flags in nearby pixels.  This was discovered to be due to charge blooming.
Charge blooming~\cite{Zengilowski2021} is related to the brighter-fatter effect~\cite{Plazas_2018}, a known detector effect causing brighter stars to appear broader due to charge repulsion from previously collected charges.  Similarly, charge blooming occurs when charge continues to be photo-generated at a saturated pixel that cannot accept additional charge. Instead, the excess charge is collected by its nearest neighbor pixels, causing the charge in these neighboring pixels to accumulate faster in subsequent frames.  When the wells of these immediate neighbors fill, the non-saturated nearest neighbors of those pixels begin to collect the excess charge and so on, causing charge to effectively bloom outward from the source.  This continues until the wells are full, the illumination changes, or the detector is reset.  

Stars sufficiently bright to saturate the central pixel will trigger the OVERFLOW flag, and excess charge will bloom outwards to nearest neighbor pixels.  
The change in slope that occurs in response to charge bloom trips the TRANSIENT flag in SPHEREx's SUR software algorithm.  This causes pixels near the center of bright stars to trigger a TRANSIENT flag. Therefore, OVERFLOW flags are often surrounded by one or more TRANSIENT flags. 

With the OVERFLOW flag triggered at roughly half of the full well and a total of 77 frames in the SUR, we calculate that the maximum current in the absence of noise for which the OVERFLOW flag is \textit{not} triggered is approximately 630~$e^-/$s or 470~$e^-/$s for the SWIR or MWIR detectors, respectively.  Saturating stars will have signals greater than twice this value.  Early overflow pixels reach the OVERFLOW threshold within the minimum required five samples and thus have a minimum signal of 12,000~$e^-/$s or 9,000~$e^-/$s respectively for the SWIR or MWIR detectors.  Pixels brighter than this threshold will be assigned NaN in the flux image.  With the level of noise in the measurement on the order of 0.1~$e^-/$s, the signals of these saturating stars are orders of magnitude larger than the background signals of interest.\cite{Bock_2026}

\begin{figure}[h]
\begin{center}
\begin{tabular}{c}
\includegraphics[width=0.95\textwidth]{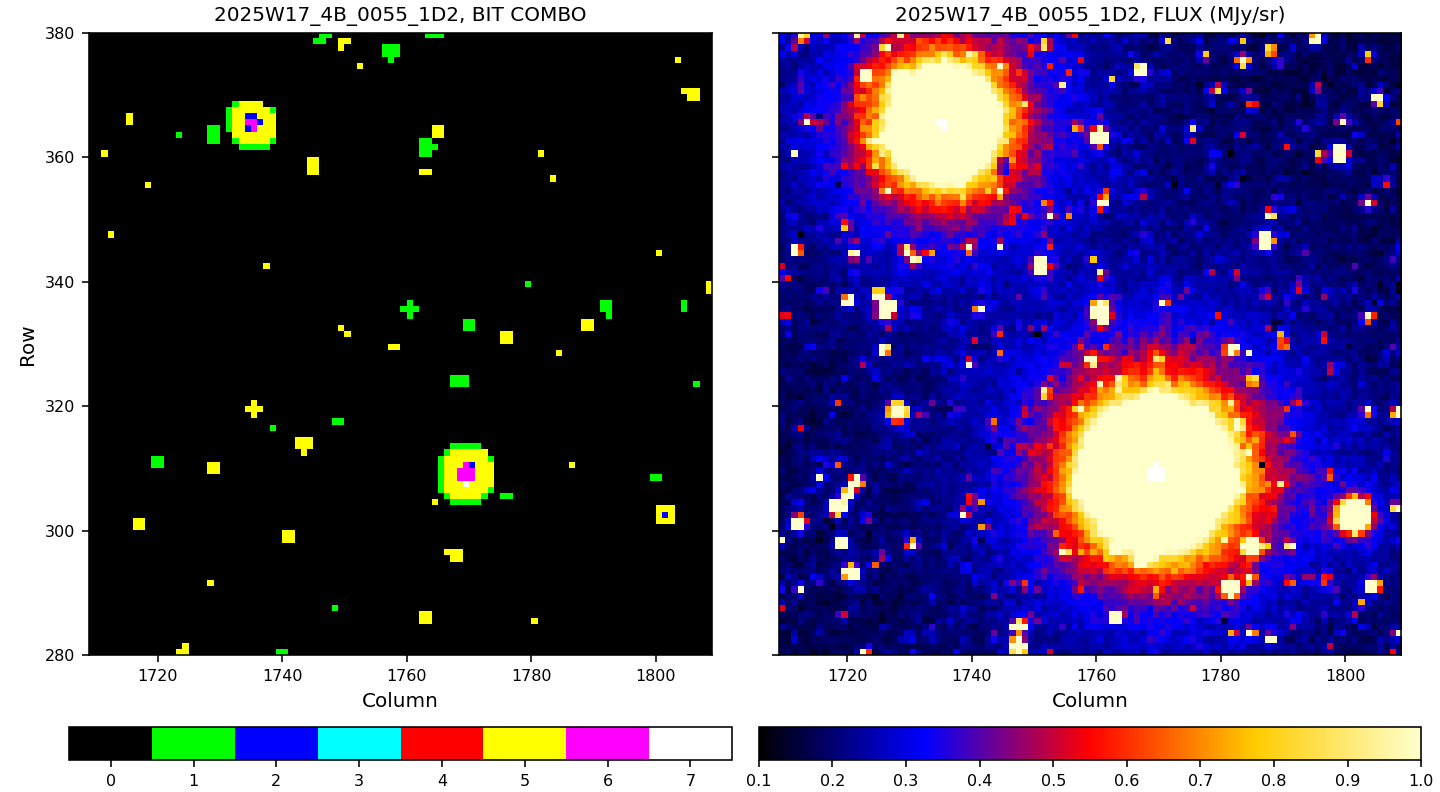}  
\end{tabular}
\end{center}
\caption 
{ \label{fig:bloomexamples}
\textbf{\textit{Source Bloom Flag Combinations.}} \textit{Left:} This image shows three different flag combinations that accompany charge bloom from different saturating astrophysical sources.  The brightest source has the largest core of overflow pixels, most which reached the overflow threshold before transient detection began, resulting in early overflow bit combinations.  All are surrounded by an early transient ring (yellow) and the largest sources also have a late transient ring (green).  Transients associated with cosmic rays can also be seen in the small clusters of yellow or green pixels indicating early or late detection respectively.  \textit{Right:} This image shows the measured flux of this region in MJy/sr.  The three saturating sources have large pale yellow regions with white NaN centers where the early overflow pixels are located.  Other non-saturating sources have smaller pale colored regions and cosmic ray hits occurring early in the exposure can manifest as dark regions, such as the block to the lower right of the source on the upper left, either due to noisy flux measurements or truncation of the SUR before charge bloomed to that radius, impacting the measured flux in surrounding pixels.  Areas outside the transient-flagged region may still be impacted by charge bloom with charge deposits lower than the transient threshold.}
\end{figure}

Very bright stars will often have a collection of early overflow pixels surrounded by a wide ring of early transients.  This ring often displays a narrow border of late transients.  
An example of three combinations resulting from saturating stars is shown in Fig.~\ref{fig:bloomexamples}.
These flagging combinations are typically circular in shape and coincide with a significant broadening to the observed extended point spread function (PSF).
In contrast, the faintest saturating stars can be indicated by a single overflow flagged pixel next to a single late transient.  In these cases, it is likely that other neighboring pixels also receive an influx of additional charge below the threshold necessary to trigger transient detection.  
Since the flux reported in pixels with these flag combinations is measured prior to the triggering event, it can be indicative of the presence of a source.  Therefore, circular pixel groupings with bit combinations as illustrated in Fig~\ref{fig:bloomexamples} that display measured flux commensurate with a source are assigned the BLOOM flag to distinguish them from other transient events.

\subsection{Extended Blooming Effects}
\label{ss:bloomingeffects}

\begin{figure}[h]
\begin{center}
\begin{tabular}{c}
\includegraphics[width=0.95\textwidth]{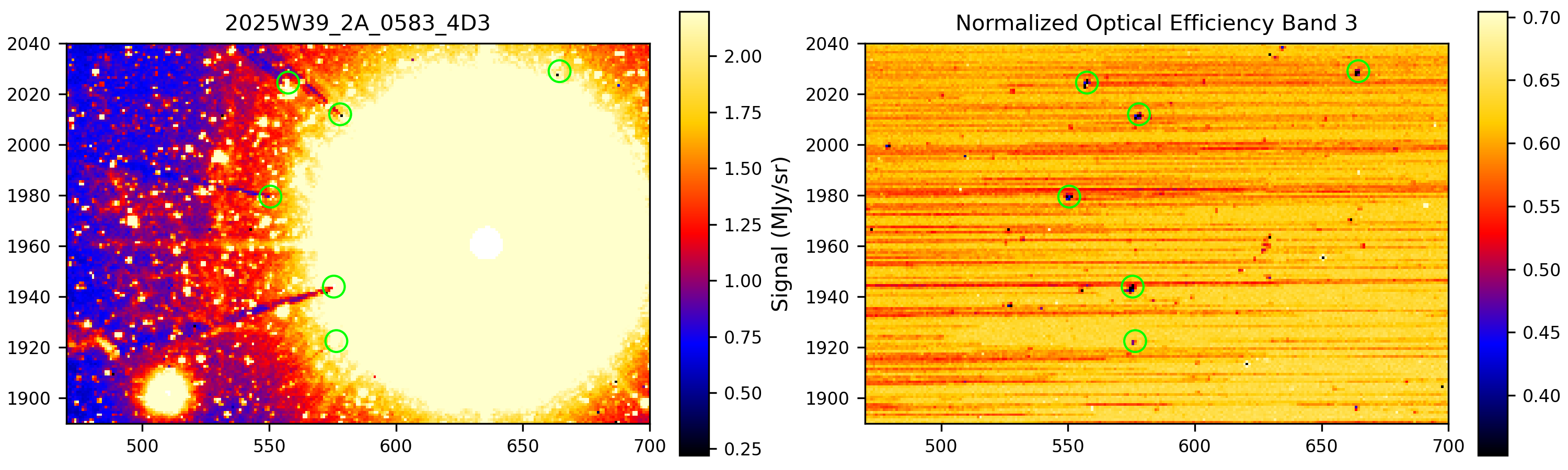}  
\end{tabular}
\end{center}
\caption 
{ \label{fig:bloomshadows}
\textbf{Bloom shadows originate at low efficiency groupings of pixels.} The left-hand panel illustrates how groupings of low efficiency pixels (lime circles) reduce the spreading of blooming signal radially beyond their location.  The right-hand panel shows the optical efficiency of those same pixel groupings normalized to the 99.9\textsuperscript{th} percentile of measured optical efficiency on the array.  These low efficiency pixel regions are the likely sources of the observed shadows.
} 
\end{figure}

For the brightest stars and the occasional planet within the field of view, excess signal can be observed tens to hundreds of pixels from the central transient cluster that we flag for blooming.  We assert that this extended region of excess signal originates within the detector itself, incited by the excessive flux in the center of the bright source.  This is supported by the geometry of its features.  For example, when the bright source is in proximity to groupings of known bad pixels one can sometimes observe an apparent wedge-shaped radial ``shadow" that manifests beyond them, as illustrated in Fig~\ref{fig:bloomshadows}. 
The wide reach of this signal and resulting shadows may point to the observed bloom's source being photo-emissive in nature, originating at the center of the source where photon flux pushes the detector diode into forward bias as it continues to generate charges after the well is full.~\cite{Zengilowski2021} We speculate that the forward-biased detector diode emits like an LED at its cutoff wavelength.
This light then propagates radially, trapped within the HgCdTe material like a light-pipe.  The efficiency of photon detection at the detector's cutoff wavelength is low, giving the photons a long mean-free-path.  As observed thus far, only groups of pixels with low efficiency coupled with high dark current or electronic drift result in a visible reduction in flux.  Reasonably, damage to the HgCdTe could cause these regions, affecting the ease of photon propagation and resulting in absorption.  One can also see a horizontal bright ``ray" in Fig~\ref{fig:bloomshadows} that looks like a diffraction spike.  This feature is only seen on blooming sources in both horizontal and vertical directions aligning with the detector readout.  Following the forward-biased pixel glow explanation, the horizontal path between pixels may provide an unobstructed path for these photons to follow, causing the observed excess signal.

\begin{figure}[h]
\begin{center}
\begin{tabular}{c}
\includegraphics[width=0.90\textwidth]{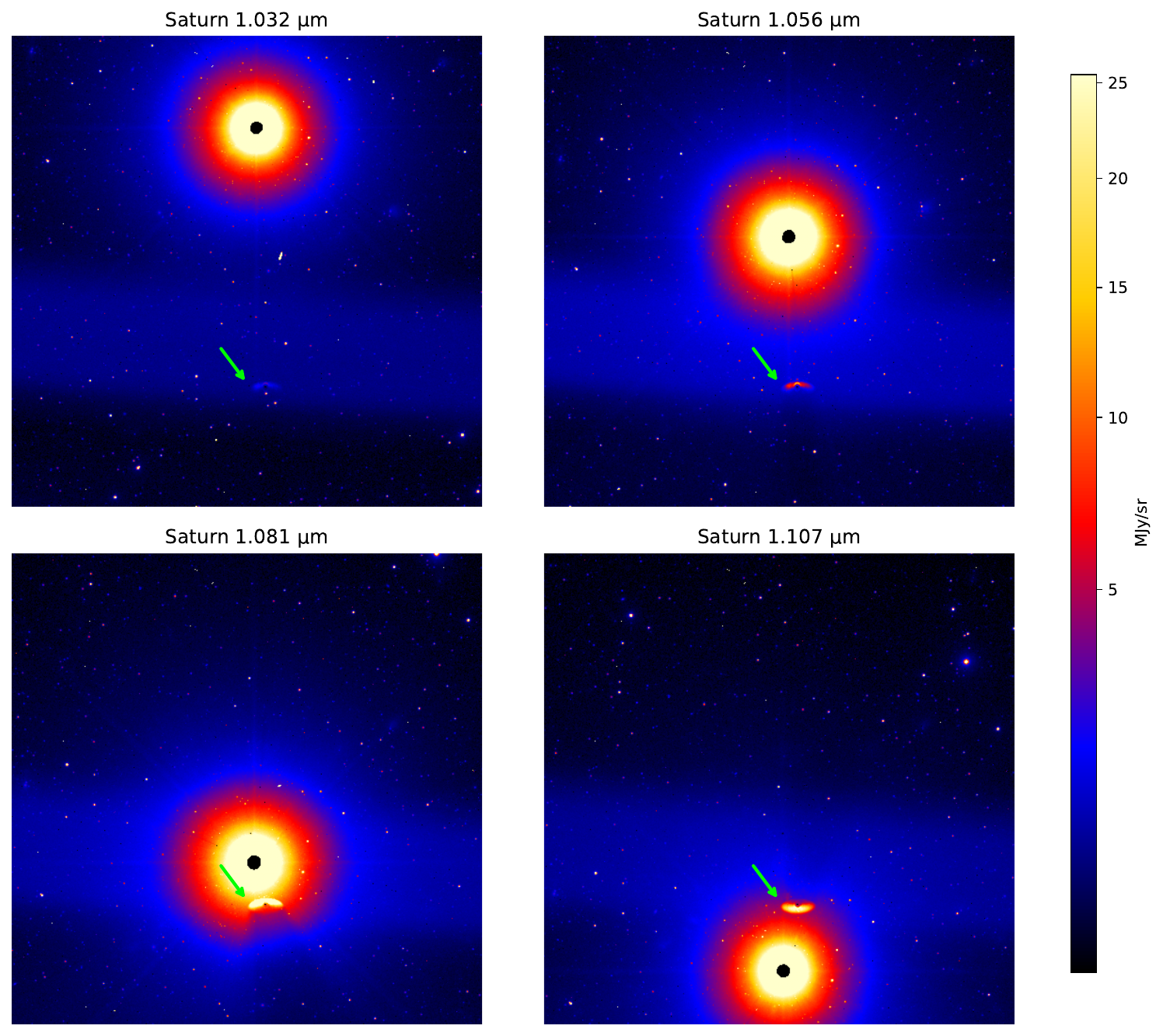}  
\end{tabular}
\end{center}
\caption 
{ \label{fig:macaroni}
\textbf{Images of Saturn and associated bad pixel blooming feature.} As the telescope slews to different positions, an image of Saturn is captured at different wavelengths passing through the Helium airglow spectral feature~\cite{hui2026observationsatmosphericheliumoxygen}, shown here with square-root scaling.  Identified by a lime green arrow, a group of bad pixels disturb the blooming signal from Saturn, causing a crescent-shaped feature of excess current to appear on one side of the group, with a ``shadow" of lesser current radiating outwards from the source on the far side.  The top-left panel demonstrates the unexpectedly large extent of excess signal accompanying bloom - over 270 pixels.  
} 
\end{figure} 

Planets are excessively bright compared to the stars in typical SPHEREx observations, and thus probe the detector's blooming behavior in the 
rare occurrence of the extreme regime. In images of the planet Saturn on Band 1 shown in Fig~\ref{fig:macaroni}, we observe a crescent-shaped build up of charge on the side of a bad pixel closest to the source.  The pixel grouping that causes this phenomenon is located in the He 1.083~$\mu$m line on Band 1~\cite{hui2026observationsatmosphericheliumoxygen}. 
The region that shows excess charge collection around the bad pixel grouping has a higher measured optical efficiency than the local average.  The forward-biased pixel glow explanation could point to a longer cutoff wavelength in this localized region, enhancing collection of these long wavelength photons and resulting in the observed excess signal and the shadow beyond.

While the theory of forward-biased pixel glow has not been tested in the lab, it explains the observed phenomena, including the extended reach and interactions with the low-efficiency bad-pixel regions.  If an accurate explanation, this phenomena warrants further study to mitigate its contamination of low-background astronomical images where saturating signals and hence blooming sources are present.
To mitigate this effect in SPHEREx data, we have increased the size of the source mask for the brightest sources to compensate for the additional electronically-induced charge at large radii.  However, a non-sidereal source like Saturn that is not included in the source catalog may not have an associated mask that captures the full extent of the observed blooming effects.

\section{Snowballs and High Energy Cosmic Rays}
\label{s:snowballs}

\begin{figure}[h!]
\begin{center}
\begin{tabular}{c}
\includegraphics[width=0.95\textwidth]{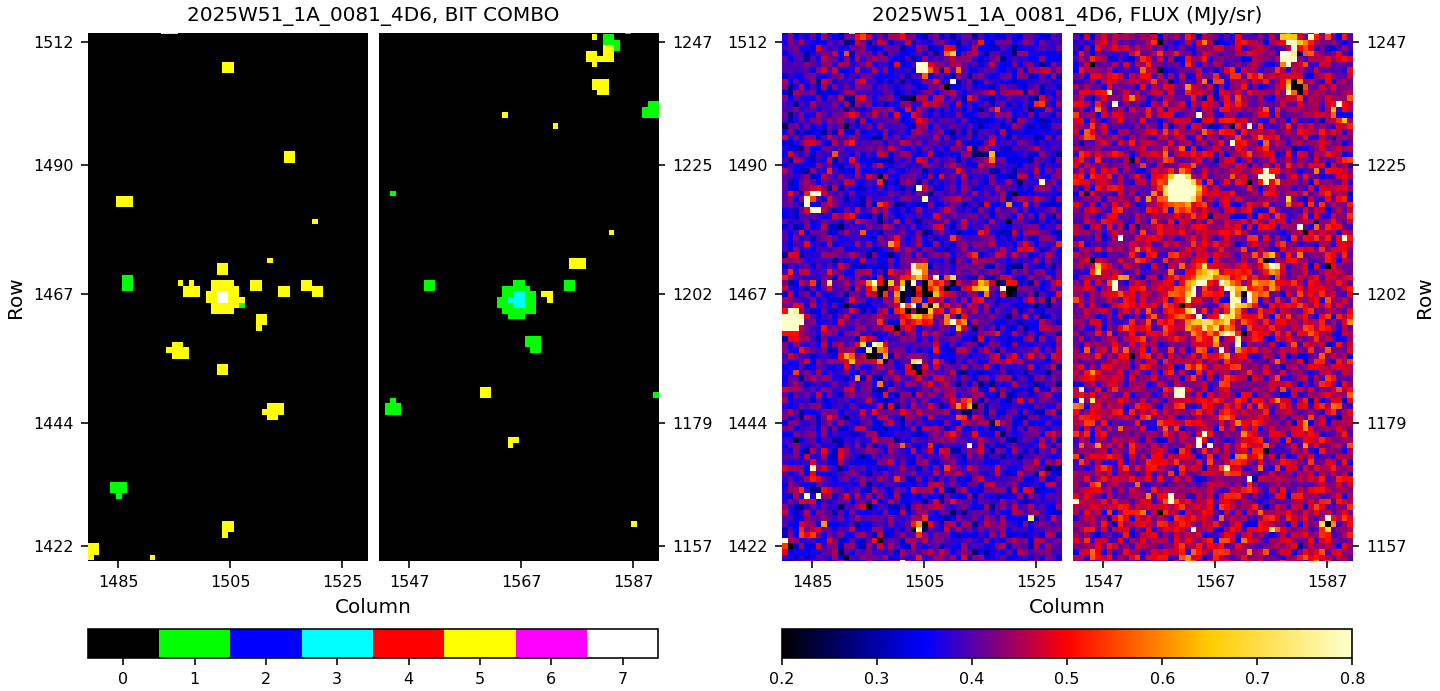}  
\end{tabular}
\end{center}
\caption 
{ \label{fig:snowexamples}
\textbf{Snowball flag combinations.} \textit{Left:} This split-panel image shows two different flag combinations that are associated with high energy cosmic ray hits or snowballs.   The left (right) split panel shows early (late) hits with bit combinations summing to inner:outer relationships of 7:5 (3:1) respectively.  The left-hand split panel also shows the common occurrence of transient clusters near the impact that share the same timing (early) as the snowball source. \textit{Right:} This split-panel image shows the measured flux of these regions in MJy/sr.  It is clear that the measured flux in the pixels where the transient was observed is low, indicating that these are particle detections and not astrophysical sources. The excess charge surrounding these impacts appears as a halo shape surrounding the low flux center.
} 
\end{figure}

In contrast with the significant flux measurements observed with blooming events, a low flux measurement is likely due to the lack of a prior flux source, pointing to a cosmic ray hit or high energy particle detection, the latter of which can be sourced by a cosmic particle or radioactive decay within the detector and has been termed a snowball~\cite{Green2020Snowballs, birkmann2022inflightnoiseperformancejwstnirspec}.  
These particle detections can be sufficiently energetic to trigger the OVERFLOW flag and generate signal in neighboring pixels similar to blooming.  We therefore look for combinations of OVERFLOW and TRANSIENT flags pointing to this type of event, such as those in Fig~\ref{fig:snowexamples}, after removing blooming events from consideration.

Truly a transient event, particle detection results in a single contribution of charge somewhere along the integration ramp, triggering the TRANSIENT flag.  Thus, unless they overlap with a source they typically have low recorded flux. Larger energy particles can affect multiple pixels simultaneously, saturating some and either blooming or otherwise depositing partial charge in other neighboring pixels.  This behavior can be (but is not always) circular, and the apparent bloomed charge region (transient ring surrounding overflow-transient core) has more ragged edges than similar clusters resulting from bright sources.   
Both the central core and surrounding TRANSIENT ring share the same state of the SUR\_ERROR flag, distinguishing it from blooming flag combinations in which the state of this flag is often mixed.  Detections such as these with large central transient-overflow regions are likely due to high energy particles.  All particles detected in this manner with transient-overflow cores and surrounding transient rings are assigned the SNOWBALL flag, regardless of the area of the array affected.  Thus, higher energy cosmic rays are included in this group.

\subsection{Snowball and Energetic Cosmic Ray Halos}
\label{ss:crhalos}

In addition to the group of transient-flagged pixels observed in the flagging layer, increases in signal are observed in neighboring pixels due to interpixel capacitance (IPC), blooming, or charge depositions below the transient flagging threshold. While low-energy particle detections that do not 
\begin{figure}[h]
\begin{center}
\begin{tabular}{c}
\includegraphics[width=0.95\textwidth]{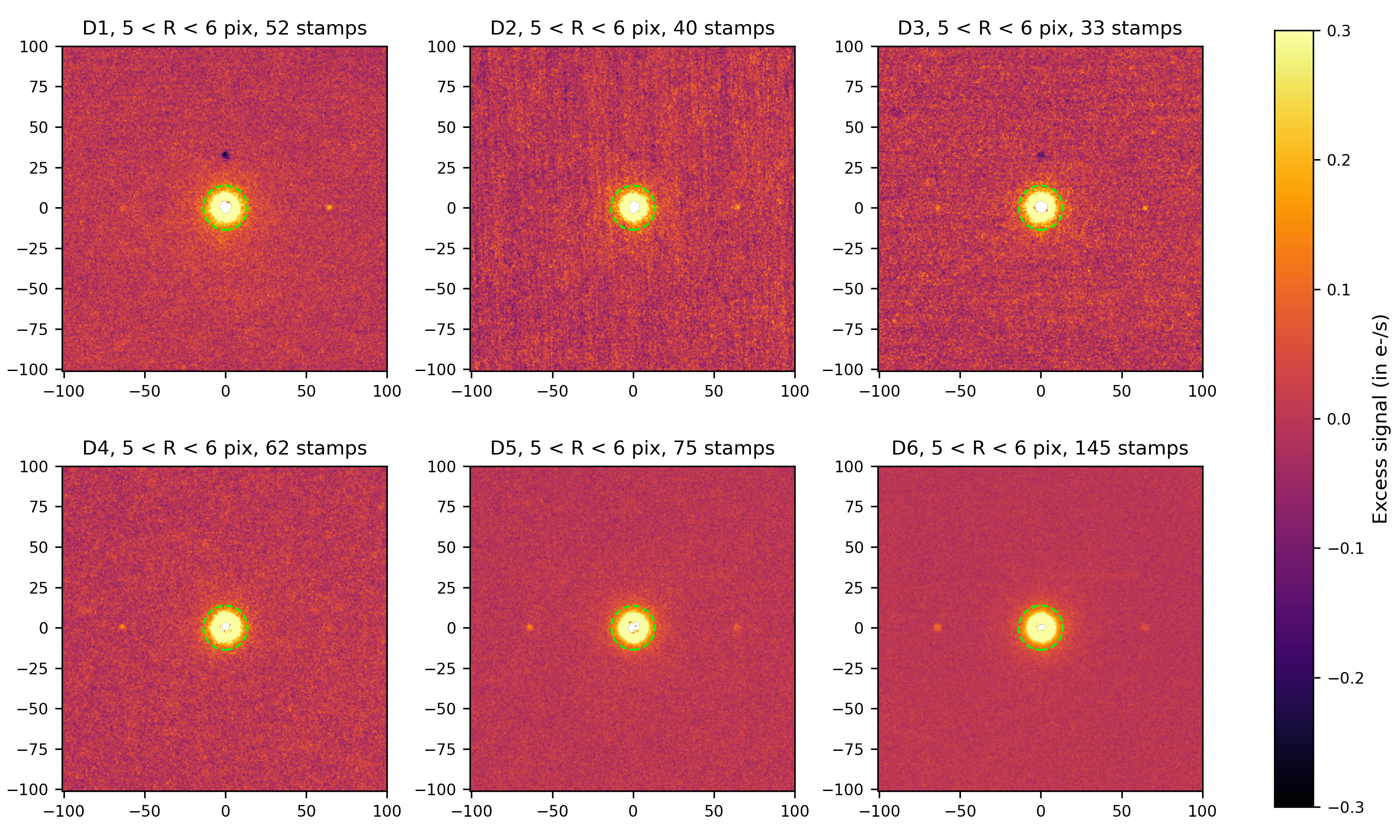}  
\end{tabular}
\end{center}
\caption 
{ \label{fig:stackCRHaloResiduals}
\textbf{Finely gridded snowball stack for similarly sized snowballs by detector.} SWIR bands and MWIR bands shown in the top three and bottom three panels respectively show the sigma-clipped median excess signal surrounding the transient region for a finely gridded and centered stack of snowballs exhibiting a transient region with an average radius between 5 and 6 pixels.  The number of stamps included in each stack is given in the panel title. 
Snowballs of all sizes are more common in the MWIR detectors, with detector 6 being the most susceptible.  Well depths of the MWIR detectors are smaller than that of the SWIR detectors, thus larger snowball radii may result for the same size energy deposit.  One can also see crosstalk types 1 and 2 (discussed in Section~\ref{s:crosstalk}) present in these stacks.
} 
\end{figure} 
trigger the OVERFLOW flag largely affect their nearest neighbors, snowballs can have a wide-reaching effect on observed signals. To prevent this extended signal from contaminating data, we have called this excess signal the snowball ``halo'' and have assigned a flag for this as well.

\begin{figure}[h]
\begin{center}
\begin{tabular}{c}
\includegraphics[width=0.95\textwidth]{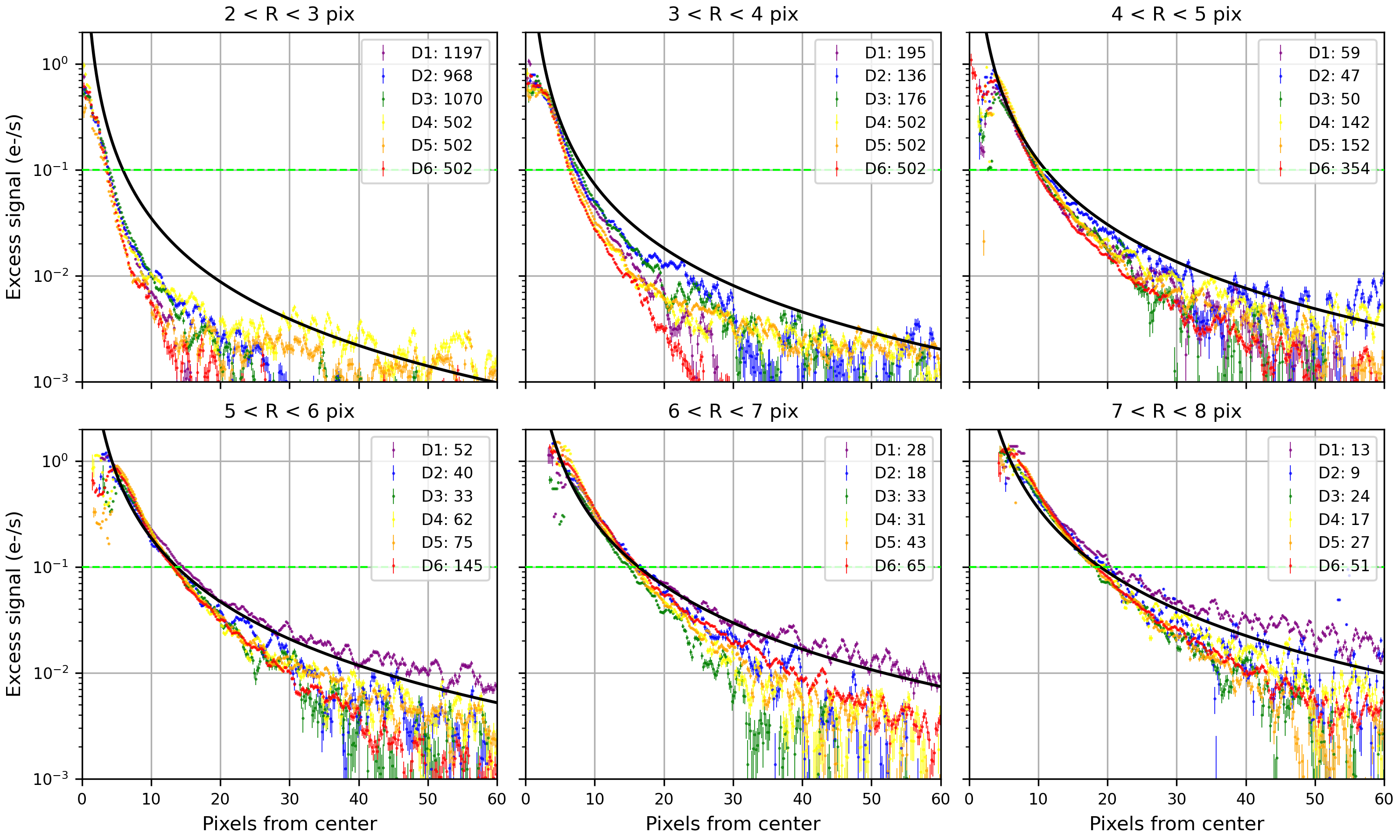}  
\end{tabular}
\end{center}
\caption 
{ \label{fig:stackCRHaloResidualsRadial}
\textbf{Radial average for selected bins of snowball radii by detector.} Plotted in each panel is the sigma-clipped mean signal by radius for snowballs with average radii within the selected bin.  The legend illustrates points with error bars representing the standard error in each measurement for each detector followed by the number of stamps used in determining the radial average.  Behavior is similar between detectors and is reasonably approximated by the empirical model in Equation~\ref{eq:snowhalo} (black line) up to the desired masking level for a range of sizes.  The smaller radii snowballs have a much smaller radial reach, likely due to the lower energy deposition remaining more localized.
The model is used to set the halo mask bit for these snowballs to $0.1~e^-/s$, illustrated by the lime green line. 
}
\end{figure}

To determine the extent of the snowball halo, we stacked the signals of snowballs of similar size, which was determined by the number of pixels in the transient cluster.  We then masked known sources and quantified the excess signal by the sigma-clipped radial average past the edge of the TRANSIENT-flagged region.  
Fig~\ref{fig:stackCRHaloResiduals} shows finely gridded stacks of snowballs of similar size by detector, with Fig~\ref{fig:stackCRHaloResidualsRadial} illustrating the sigma-clipped mean radial profile for selected radial bins for each detector.  
These figures illustrate the similarity in behavior across the detectors for the same size snowball, with the largest deviations being variations in signal at larger radii, possibly due to background variations in each detector. 

The empirical formula chosen to describe the excess current 
($i_{halo}$ in $e^-/s$) as a function of radius, $r$, is given by
\begin{equation}
\label{eq:snowhalo}
i_{halo}(r)= \frac{0.2 A_{snow}}{r^2},
\end{equation}
where $A_{snow}$ is the number of pixels in the transient cluster.  
This formula closely follows the radial behavior at lower radii for all six detectors, though it over-predicts for the smallest measured snowball sizes.  However, for ease of implementation and a conservative approach, we applied the model uniformly. 
Since $0.1~e^-/s$ is on the order of noise, this was chosen as the lower threshold of excess current out to which the snowball halos should be masked.  
Fig~\ref{fig:stackCRHaloResiduals} includes a dashed lime circle indicating the radius to which the halo would be masked for each detector.  Similarly, Fig~\ref{fig:stackCRHaloResidualsRadial} shows a lime line at 0.1~$e^-$/s corresponding to the masking threshold.

\begin{figure}[h]
\begin{center}
\begin{tabular}{c}
    \includegraphics[width=0.95\textwidth]{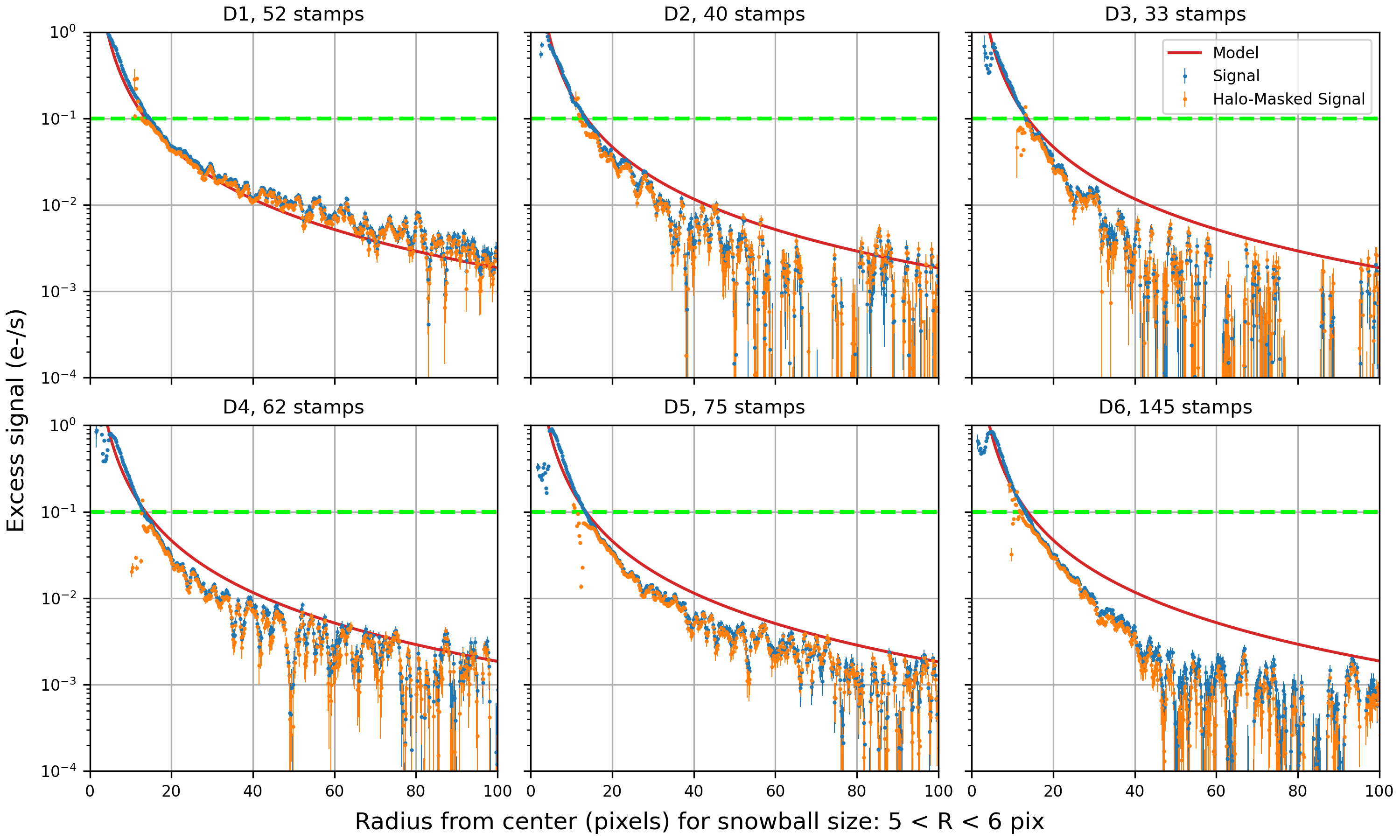}  
\end{tabular}
\end{center}
\caption 
{ \label{fig:stackCRHaloResidualsRadial67}
\textbf{Excess signal and halo-masked residual by detector.} SWIR bands and MWIR bands shown in the top three and bottom three panels respectively show the radial average of the sigma-clipped median excess signal surrounding the transient region for a finely gridded and centered stack of snowballs exhibiting a transient region with an average radius between 5 and 6 pixels.  The number of stamps included in each stack is given in the panel title.  Shown in orange is the same calculation with the halo-masked pixels excluded.  Plots illustrate the excess signal is effectively reduced with the halo mask.
} 
\end{figure}

Some images can have no snowballs or high-energy cosmic ray hits while others can have many.  We find that as the transient fraction increases following periods of solar activity~\cite{Nguyen2026}, so do the occurrences of these high-energy events. On average, pixel groups flagged as snowballs contribute approximately 800 or 4000 pixels to the halo mask for the SWIR or MWIR detectors, respectively.  
The combination of nearest neighbor flagging of transients and snowball halo masking results in the HALO mask bit in the flagging layer of SPHEREx L2 data.  The application of the halo mask is illustrated in Fig~\ref{fig:stackCRHaloResidualsRadial67}, 
which compares the sigma-clipped radial average of the same stamps in Fig~\ref{fig:stackCRHaloResiduals} with (blue) and without (orange) masking the pixels flagged for HALO.  This figure, which incorporates the HALO flag in the L2 flag layer, demonstrates that halo masking effectively removes pixels with excess signal from science analysis.

\section{Crosstalk}
\label{s:crosstalk}
\begin{figure}[hb]
\begin{center}
\begin{tabular}{c}
\includegraphics[width=0.95\textwidth]{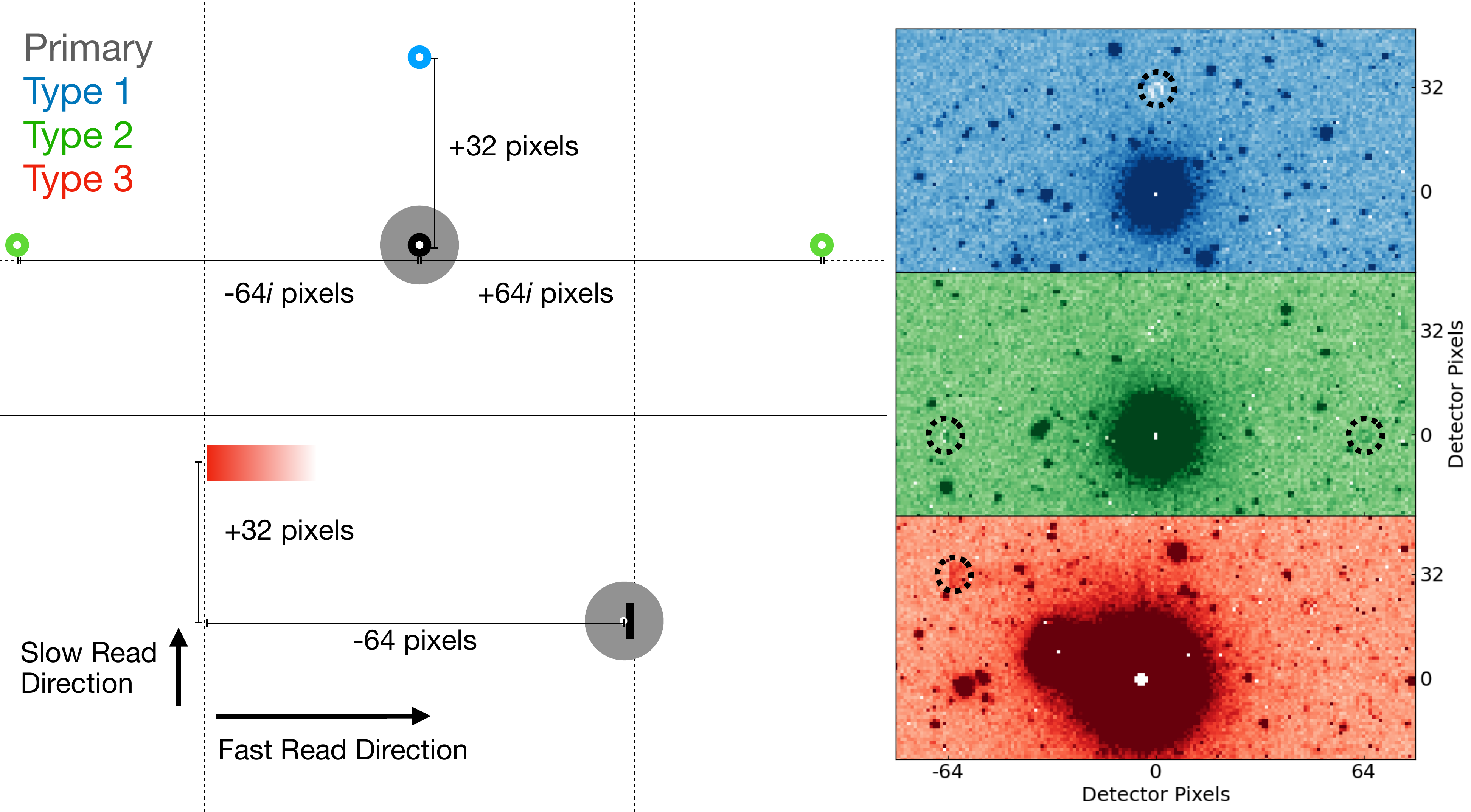}  
\end{tabular}
\end{center}
\caption 
{ \label{fig:crosstalkexamples}
\textbf{Overview of electrical crosstalk effects in the SPHEREx H2RG detectors.} We observe three types of electrical crosstalk that are present when bright sources illuminate the detector array.  Bright sources in this illustration are shown in ``negative" form, where more flux is rendered in a darker tone.  At the center of these bright regions are early overflow pixels with no determinable flux measurement, and so are assigned NaN and thus appear white.  Pixels with unusually low or negative flux may also appear white on this scale.
Electrical crosstalk Type 1 (blue) is an image of the large photocurrent pixels (dark circle) around a bright source (saturated pixels displayed in white at the center of the dark circle) displaced by 32 pixels in the slow read (row-to-row) direction, consistent with the NSKIP parameter of the row-chopping technique. Electrical crosstalk Type 2 (green) generates repeated sources at multiples of 64 pixels in both the positive and negative directions in the fast read (column-to-column) direction.  This is thought to be due to true electrical crosstalk, as these pixels are read simultaneously. Electrical crosstalk Type 3 (red) is a smeared signal, but only occurs for bright (but not saturated) pixels at the edge of a channel. We suspect this may be due to some kind of hysteresis sensitive to large signal step functions in the Video 8 readout system. The right-hand panels show examples of each type of crosstalk color coded as in the schematic, with the incident stars being HD88230, HD87357, and HD233705. Dotted circles highlight the crosstalk images in each case.} 
\end{figure}

Fig~\ref{fig:crosstalkexamples} illustrates three types of electrical crosstalk observed in SPHEREx data that manifest when bright sources illuminate the detector array.  In all cases, the crosstalk image depends on the photocurrent in a pixel rather than the total integrated charge, meaning that saturated pixels do not generate a strong crosstalk image. 
As discussed in Section~\ref{s:blooming}, after the central pixels of a bright source saturate, the signal blooms out to neighboring pixels, increasing their measured current. As a result, true crosstalk images (Type 1 and 2) often appear as a small ``donut” generated most strongly by non-saturated pixels in the blooming region surrounding the central saturated pixels in images of magnitude $6 {-} 9$ sources.  The three types of crosstalk are discussed in the following subsections in order of decreasing severity.

\subsection{Type 1 Crosstalk}
\label{ss:xt1}

The unique noise-suppression readout technique of row-chopping employed by SPHEREx is spatially connected to the appearance of Type 1 crosstalk.  Row-chopping~\cite{Heaton_2023} is 
a novel readout scheme employed by SPHEREx 
in which each fast row read is separated by NSKIP (a parameter set to 32) rows from the previous row.  This continues until the end of the array is reached, whereupon the readout returns to the next row in the sequence and continues this pattern.  SPHEREx uses this technique to mitigate $1/f$ noise by modulating the low frequency drift to a higher frequency to minimize its impact.  Interspersed with these row readouts are samples of the voltages on the VIDEO 8 boards.\cite{Nguyen2025}  These samples are called ``phantom" pixels and are used to further mitigate noise by sampling the voltage drift of the electronics.

\begin{figure}[h]
\begin{center}
\begin{tabular}{c}
\includegraphics[width=0.95\textwidth]{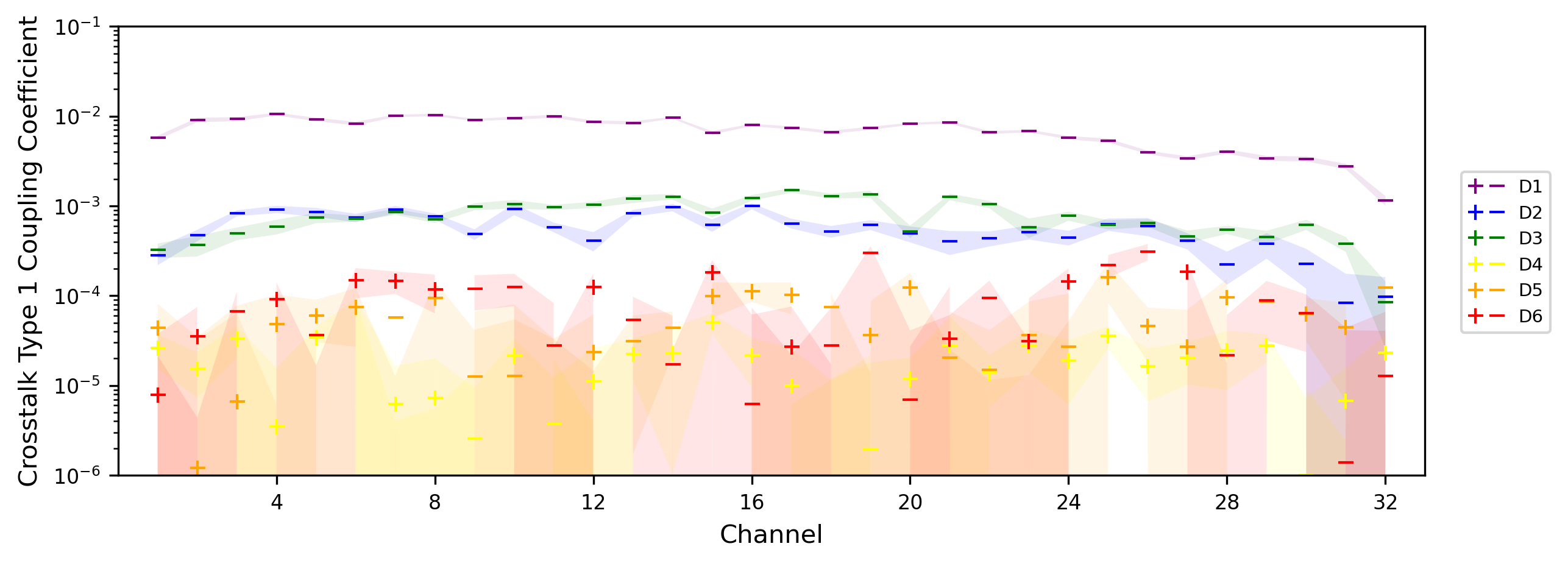}  
\end{tabular}
\end{center}
\caption 
{ \label{fig:type1xtplot}
\textbf{Type 1 crosstalk coupling coefficient.} Shown in this figure is the coupling coefficient between the signal and its apparent crosstalk image.  Two symbols are used for each band and channel, the $+$ and $-$ symbols indicating that the coupling produces a positive and negative signal respectively.  Standard error is shown as the shaded region around data points.  Bands 1-3 have observable crosstalk of this flavor that appears only as a negative imprint.  Parameters for Bands 4-6 are consistent with noise, producing both positive and negative values about zero.  By far, the largest coupling can be seen in the negative imprint observed in Band 1.} 
\end{figure}

If a low signal region follows a fast row read that includes a bright source, a negative imprint of the source can sometimes be seen exactly 32 pixels above the source.  This is consistent with the row-chopping scheme and has also been seen in CIBER\nobreakdash-2 data with a spacing of 25 pixels, consistent with its row skip parameter.~\cite{Fazar_rowchop}  This is illustrated as a white circle indicating negative signal directly above the strong signal (dark) region at the top of Fig~\ref{fig:crosstalkexamples}.  The coupling between the source and its negative imprint is band-dependent, as illustrated in Fig~\ref{fig:type1xtplot}, which shows the coupling coefficient of the negative imprint as a $-$ symbol.  The largest coupling is observed in Band 1.  Bands 4, 5 and 6 exhibit parameters consistent with noise, meaning there is no evidence for this kind of crosstalk operating in these arrays.

\subsection{Type 2 Crosstalk}
\label{ss:xt2}
\begin{figure}[h]
\begin{center}
\begin{tabular}{c}
\includegraphics[width=0.95\textwidth]{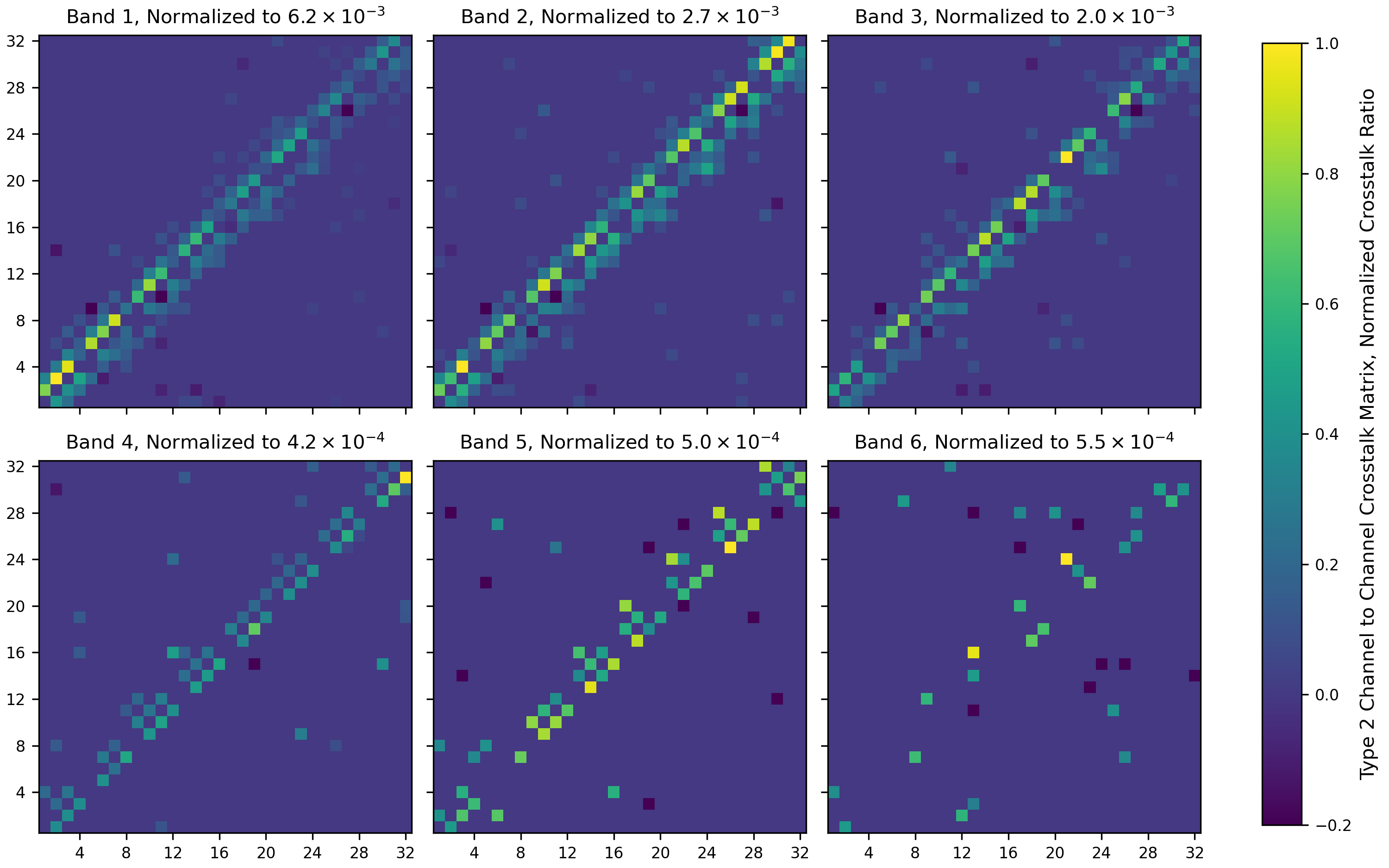}  
\end{tabular}
\end{center}
\caption 
{ \label{fig:type2xtmatrices}
\textbf{Type 2 crosstalk coupling coefficient matrices.} Each panel in this figure shows the crosstalk coupling coefficient between channels. The vertical channel is the channel containing the source and the horizontal channel is the channel to which the signal couples. Each channel and detector has unique behaviors, with Band 1 exhibiting the strongest crosstalk of this type.  To demonstrate the variation between channels and detectors, each panel is normalized to the highest observed coefficient and demonstrates similar channel-to-channel relationships.  Shown at the top of each matrix is the normalization factor, which is tabulated to more significant figures in Table~\ref{tab:maxtype2xtcoeff}, illustrating the overall strength of the coupling.} 
\end{figure}

Type 2 crosstalk is produced when a bright star generates large photocurrent, which then blooms out to its nearest neighbors.  Pixels in neighboring channels that are read out at the same time as these high current pixels can display excess charge mimicking the shape of the bright region.  This effect is most sensitive in the nearest neighboring channels, though can sometimes be seen many channels away.  The strongest coupling is on the order of $\sim 10^{-3}$ and is thought to be due to true electrical crosstalk.  Fig~\ref{fig:type2xtmatrices} shows the channel-to-channel coupling coefficients, normalized to the maximum observed coupling, which is tabulated in Table~\ref{tab:maxtype2xtcoeff}.

\begin{table}[h]
    \centering
    \begin{tabular}{c|c}
        BAND & MAX COUPLING COEFFICIENT \\
        \hline
        1 & 0.006194 \\
        2 & 0.002651 \\
        3 & 0.002027 \\
        4 & 0.000424 \\
        5 & 0.000500 \\
        6 & 0.000550 \\
        
    \end{tabular}
    \vspace{0.1cm}
    \caption{
    Maximum coupling coefficient of type 2 crosstalk used to normalize the matrices in Fig~\ref{fig:type2xtmatrices}.  The strongest crosstalk is once again observed in Band 1 and the SWIR bands (Bands 1-3) couple by approximately an order of magnitude more than the MWIR bands (Bands 4-6).}
    \label{tab:maxtype2xtcoeff}
\end{table}

\subsection{Type 3 Crosstalk}
\label{ss:xt3}
\begin{figure}[h]
\begin{center}
\begin{tabular}{c}
\includegraphics[width=0.95\textwidth]{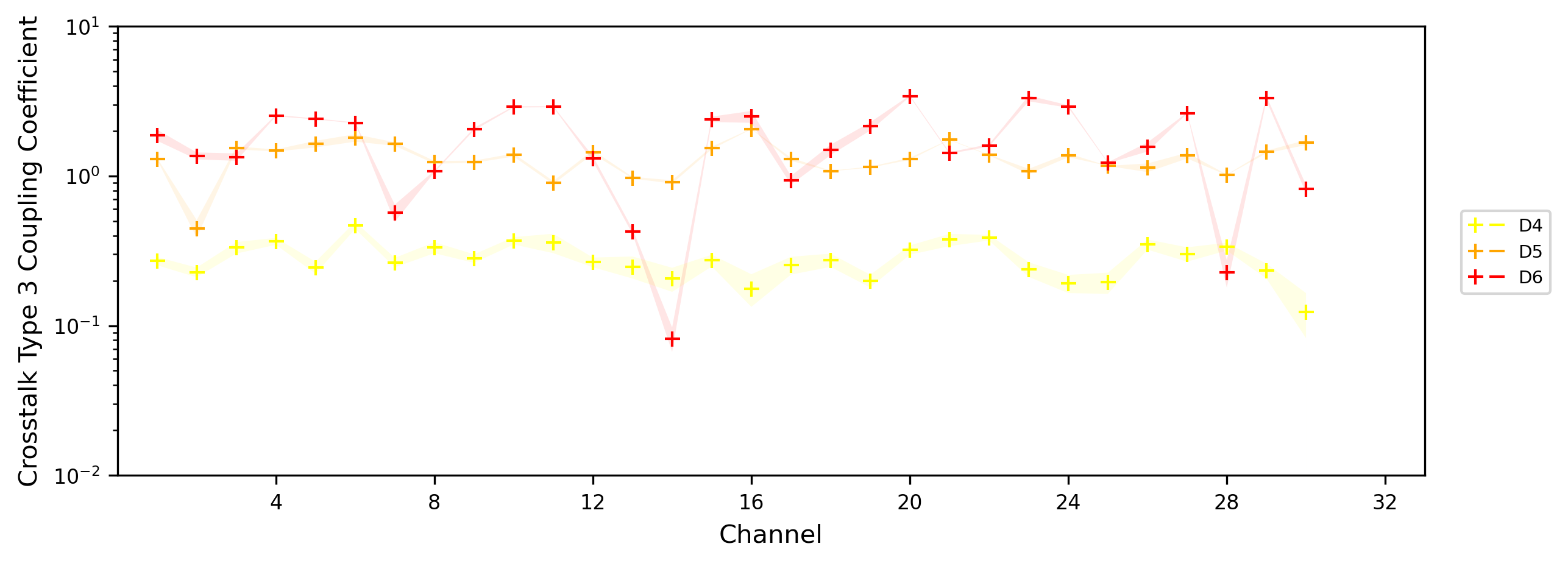}  
\end{tabular}
\end{center}
\caption 
{ \label{fig:type3xtplot}
\textbf{Type 3 crosstalk coupling coefficient.} 
Shown in this figure is the coupling coefficient between the source and the observed smear crosstalk image.  Note that only Bands 4 through 6 are captured here and that this crosstalk always generates a positive signal, indicated by the positive symbol, $+$.  The vertical spread of the shaded region connecting neighboring channels illustrates the standard error in these measurements. The coupling coefficient is strongest in Band 6 and can produce a total excess signal greater than the pixel source.  This effect is likely due to some interaction with the Video 8 voltages in the readout chain, and is not true crosstalk.} 
\end{figure}

Type 3 crosstalk results from a bright source which lands on the edge of a channel such that the bright pixels are last pixels in the row.  Due to the readout scheme, the next read is a phantom pixel followed by the first detector pixel in a row located 32 rows away.  Under certain electronic conditions this can cause a bright ``smear" in the signal along the subsequently read row.  The profile of the smear is empirically represented by
\begin{equation}
    \label{eq:type3xtexp}
    \eta(x) = A e^{-Bx},
\end{equation}
where $x$ is the distance in pixels from the first pixel in the row of that channel, the decay rate $B$ is band specific, and $A$ normalizes the equation so that it integrates to unity over a span of 24 pixels.
The total signal encompassed in the smear for type 3 crosstalk is then given by the integral of the signal across the first 24 row pixels in that channel,
\begin{equation}
    \label{eq:I3}
    I_3 = C\,I_s \int_{0}^{23} {\eta(x)\,dx},
\end{equation}
where $I_s$ is the signal in the detector pixel read immediately prior to the first pixel of the crosstalk smear and $C$ is the crosstalk type 3 coupling coefficient.
If the last pixel of the previous row contains a saturated pixel that has no measured current, no crosstalk will be observed in that row, causing the corresponding multi-row crosstalk smear 
to contain a row in the center displaying no type 3 crosstalk.  Analysis is limited to Bands 4-6, where this effect has been observed.  Parameters for $A$ and $B$ and the mean coupling coefficient $\bar{C}$ are given in Table~\ref{tab:abparamstype3xt}, 
\begin{table}[h]
    \centering
    \begin{tabular}{c|c c c c}
        BAND & $A$ & $B$ & $\bar{C}$ & $\sigma_{C}$ \\
        \hline
        4 & 0.1454 & 0.1302 & 0.2822 & 0.0743 \\
        5 & 0.1274 & 0.1122 & 1.3247 & 0.3154 \\
        6 & 0.0497 & 0.0144 & 1.8334 & 0.9303 \\
    \end{tabular}
    \vspace{0.1cm}
    \caption{
    Parameters used to empirically describe the smear observed in type 3 crosstalk with Equations~\ref{eq:type3xtexp} and~\ref{eq:I3}.  The parameters $\bar{C}$ and $\sigma_C$ represent the mean coupling coefficient and standard deviation respectively.}
    \label{tab:abparamstype3xt}
\end{table}
with the channel-dependence of the coupling coefficient illustrated in Fig~\ref{fig:type3xtplot}.  
We suspect this type of crosstalk is due to some effect of the sequencing in the readout chain, where the intermediate ``phantom" pixel read between the high signal pixel and the low background causes a high signal offset in the electronics that has a significant decay time. As such, it is not true crosstalk and can have a coupling coefficient greater than 1.

The empirical models presented above are used to determine the level of crosstalk signal present in affected pixels and to flag them appropriately.  For types 1 and 2, the signal in the crosstalk image is determined by multiplying the observed signal of the source by the crosstalk coupling coefficient and flagging all pixels whose crosstalk signal is predicted to exceed $0.3~e^-/s$.  For type 3, the entire subsequent row is flagged if the pixel at the end of the previously read row is sufficiently high to cause this type of excess signal, as the behavior is more loosely linked to the source signal and is subject to a variety of other factors, such as intermediate phantom pixel reads.
These crosstalk flags will appear in the flag layer of L2 data.

\section{Additional Electronic Image-Space Artifacts}
\label{s:more}

Image persistence is another known H2RG detector behavior that presents as an image-space artifact.  While not negligible in these detectors, a method of flagging pixels affected by this phenomenon has been developed~\cite{Fazar2025} and the newest model successfully incorporates charge-bloom and snowball-related impacts.  An analysis of the behavior of persistence in flight is forthcoming in (Fazar,~C. et.~al, in prep)~\cite{FazarPersistence}.

\section{Summary and Conclusions}

In summary, we described various image-space artifacts of an electronic nature in SPHEREx detectors and shared our characterization process and current masking strategies. 
Initial investigations into charge blooming were shared, starting with methods of identification in SPHEREx data.  We additionally presented effects at large radii due to excessively bright sources and interactions with low efficiency bad pixel regions.  It was asserted that this extended bloom-halo may be due to forward-bias induced pixel glow.  Further study of this behavior under extreme conditions and its association with efficiency would be beneficial for understanding the underlying causes, potentially leading to engineering solutions for future missions.  In addition, observation time and readout strategies should be optimized to minimize lengthy pixel saturation where possible.
Measurements of excess signal (termed the snowball halo) due to high energy particle detections were plotted and modeled.  The halo is effectively masked with the presented model, which minimizes its impact upon science data.  A study of these detections and their energy distributions in SPHEREx data is underway.
Three different types of observed crosstalk, two of which were discovered in flight, were characterized. Though the observed crosstalk is only evident with saturating sources and effectively masked by the presented model, further investigation of the dependence of the crosstalk coefficient upon detector parameters and set voltages would be beneficial for minimizing the effects of all three types of observed crosstalk in future missions.

Though the electronic artifacts we have discussed can be successfully managed through the masking strategies presented when they are independent and sparse, they can be much more challenging in crowded fields and during particle storms when their effects overlap.  The snowball halo and nearest-neighbor cosmic ray halo alone can increase the number of pixels affected by transients by a factor of 3 or more, leaving few good pixels remaining for science analyses.  To mitigate space environment impacts of cosmic ray showers, SPHEREx has marked affected images ($\sim2\%$ of non-SAA images) for re-observations.~\cite{Nguyen2026}  The ability to select targets to minimize data loss is extremely valuable, since the space environment is exposed regularly to such conditions.

\clearpage

\section*{End Pages}

\subsection*{Disclosures}
This author declares no conflicts of interest in publishing this information.

\subsection* {Code and Data Availability} 
Code and data may be obtained upon request.  Contact the lead author at \linkable{cmfsps@rit.edu}.

\subsection* {Acknowledgments}
We acknowledge support from the SPHEREx project under a contract from the NASA/GODDARD Space Flight Center to the California Institute of Technology (80GSFC18C0011).

Part of the research described in this paper was carried out at the Jet Propulsion Laboratory, California Institute of Technology, under a contract with the National Aeronautics and Space Administration (80NM0018D0004).

The authors acknowledge the Texas Advanced Computing Center (TACC) at The University of Texas at Austin for providing computational resources that have contributed to the research results reported within this paper. URL: http://www.tacc.utexas.edu

The authors acknowledge the discussion at the SPIE Astronomical Telescopes and Instrumentation 2026 poster session with B. Rauscher and M. D. Perrin that solidified the formulation of the theory that the forward-biased pixel glow produces photons at the cutoff wavelength of the material, propagating through the HgCdTe layer.


\bibliography{report}   
\bibliographystyle{spiejour}   

\end{document}